\documentclass[fleqn,usenatbib]{mnras}

\usepackage{newtxtext,newtxmath}

\usepackage[utf8]{inputenc}
\usepackage[T1]{fontenc}
\usepackage{ae,aecompl}

\usepackage{graphicx}	
\usepackage{amsmath}	
\usepackage{amssymb}	
\usepackage{hyperref}
\usepackage{multirow}
\usepackage{booktabs}
\usepackage{textcomp}
\hypersetup{colorlinks, linkcolor={red}, citecolor={blue}, urlcolor={magenta}}
\usepackage{morefloats}
\usepackage[separate-uncertainty,multi-part-units = single]{siunitx}
\usepackage{tabularx}

\newcommand{\ind}[2]{\ensuremath{#1_\mathrm{#2}}}
\newcommand{\Rblr}[2]{\ind{R}{#1}^{\mathrm{#2}}}

\DeclareSIUnit\solarmass{\ensuremath{M_\odot}}
\DeclareSIUnit\solarlum{\ensuremath{L_\odot}}
\DeclareSIUnit\angstrom{\ensuremath{\mathrm{\mbox{\AA}}}}
\DeclareSIUnit\parsec{\ensuremath{\mathrm{pc}}}
\DeclareSIUnit\erg{\ensuremath{\mathrm{erg}}}
\DeclareSIUnit\year{\ensuremath{\mathrm{yr}}}
\defcitealias{2015ApJ...800L..35A}{AP15} 

\title{Estimating supermassive black hole masses in AGNs using polarization of broad Mg\,II, H$\alpha$ and H$\beta$ lines}

\author[Đ.\,\,Savić et al.]{
Đorđe Savić,$^{1,2}$\thanks{E-mail: djsavic@aob.rs}
L.\,\,Č.\,\,Popović,$^{1,3}$
E.\,\,Shablovinskaya$^{4}$
 and V.\,\,L.\,\,Afanasiev$^{4}$
\\
$^{1}$Astronomical Observatory Belgrade, Volgina 7, 11060 Belgrade, Serbia\\
$^{2}$Universit\'e de Strasbourg, CNRS, Observatoire Astronomique de Strasbourg, UMR 7550, 11 rue de l’Universit\'e, F-67000 Strasbourg, \\France\\
$^{3}$Department of Astronomy, Faculty of Mathematics, University of Belgrade, Studentski trg 16, 11000 Belgrade, Serbia\\
$^{4}$Astrophysical Observatory of the Russian Academy of Sciences, Nizhnij Arkhyz, Karachaevo-Cherkesia 369167, Russia
}

\date{Received 03 February 2020. Accepted 22 July 2020.}

\pubyear{2019}

\begin{document}
\label{firstpage}
\pagerange{\pageref{firstpage}--\pageref{lastpage}}
\maketitle

\begin{abstract}
For type-1 active galactic nuclei (AGNs) for which the equatorial scattering is the dominant broad line polarization mechanism, it is possible to measure the supermassive black hole mass by tracing the Keplerian motion across the polarization plane position angle $\varphi$. So far this method has been used for 30 objects but only for H$\alpha$ emission line. We explore the possibilities this method for determining SMBH masses using polarization in broad emission lines by applying it for the first time to Mg\,II\,$\lambda$\SI{2798}{\angstrom} spectral line. We use 3D Monte Carlo radiative transfer code \textsc{stokes} for simultaneous modeling of equatorial scattering of H$\alpha$, H$\beta$ and Mg\,II lines. We included vertical inflows and outflows in the Mg\,II broad line region (BLR). We find that polarization states of H$\alpha$ and H$\beta$ lines are almost identical and SMBH mass estimates differ by \SI{7}{\percent}. For Mg\,II line, we find that $\varphi$ exhibits an additional ``plateau'' with a constant $\varphi$, which deviates than the profiles expected for pure Keplerian motion. SMBH mass estimates using Mg\,II line are higher by up to \SI{35}{\percent} than those obtained from H$\alpha$ and H$\beta$ lines. Our model shows that for vertical inflows and outflows in the BLR that is higher or comparable to Keplerian velocity, this method can be applied as a first approximation for obtaining SMBH mass.
\end{abstract}




\section{Introduction} \label{s:intro}
Supermassive black holes (SMBHs) reside in the heart of nearly every massive galaxy in the Universe. Their mass typically range between \SIrange{d6}{d9.5}{\solarmass} \citep{1995ARA&A..33..581K}. Most of them lie dormant, but when the nearby gas is abundant, it will start the accretion process where the disk is formed. As the temperatures of the accreting matter increases, an immense amount of energy is radiated, triggering an active phase now known as an active galactic nucleus (AGNs) \citep{1964ApJ...140..796S,1964SPhD....9..246Z,1969Natur.223..690L}. Whether they are dormant or active, the gas and stars surrounding SMBHs are sensitive to their presence, allowing us to measure their mass. When in their active phase, SMBHs play an important role in shaping its environment in a process called AGN feedback \citep[][and references therein]{2012ARA&A..50..455F}. As a consequence of AGN feedback, numerous correlations of SMBH mass with the properties of the host galaxy have been found, of which the most notable is $\ind{\mathcal{M}}{bh} - \sigma_*$ relation \citep{2013ARA&A..51..511K}, implying that SMBH and the host galaxy co-evolve together \citep{2011Sci...333..182H}. Therefore, reliable SMBH mass measuring is an important task in astronomy. For that purpose, different techniques have been developed, both direct and indirect \citep[][for more details]{2014SSRv..183..253P}, with most of the methods targeting AGNs due to their high luminosity, which can be readily observed at different cosmological scales. The standard paradigm, or the so called unified model of AGNs \citep{1993ARA&A..31..473A} assumes that the SMBH is surrounded by an accretion disk which is further away from the center fragmented into an optically thick dusty torus. Dusty torus collimates the radiation in the polar direction and obscures the central region along the equatorial viewing direction. The broad line region (BLR) resides in the vicinity of the SMBH, at distances of a few to a few hundred light days, in which the gas is being photoionized by the radiation from the accretion disk. Lines are emitted due to radiative recombination and collisional excitations \citep{2013peag.book.....N} and their width of a few thousand \SI{}{\kilo\meter\per\second} is due to Keplerian motion around the SMBH \citep{1991ApJ...366...64C}. The observed dichotomy between type-1 AGNs where the broad emission lines are visible and the type-2 with only narrow emission lines in their optical spectra is largely due to an orientation effects where type-1 AGNs are observed with from close to pole-on view while type-2 AGNs are viewed at much higher inclinations, closer towards edge on view. For other AGN components and the unified model review, we refer to \citep{2015ARA&A..53..365N}.

Over the past years, the most reliable SMBH mass measurements come from the reverberation mapping of AGNs \citep{2015PASP..127...67B}. By measuring the time delay between the variability of the  ionizing continuum and the broad emission lines variability, we can obtain a photometric BLR radius. With known photometric radius, and the velocity measured directly from the broad emission line, we can obtain the SMBH mass \citep{1972ApJ...171..467B,1982ApJ...255..419B,1993PASP..105..247P}.The duration of a reverberation mapping experiment can be rather long. An individual galaxy needs to be observed over and over again for several months, while distant AGNs require even several years of successful monitoring \citep{2000ApJ...533..631K,2004ApJ...613..682P,2007ApJ...659..997K,2009NewAR..53..191S,2013ApJ...769..128B,2014ApJ...782...45D,2015ApJS..217...26B,2015ApJ...806...22D,2016ApJ...818...30S,2017ApJ...851...21G,2017FrASS...4...12I,2018ApJ...856....6D,2019ApJ...886...42D,2019ApJ...887...38G}. Hydrogen Balmer lines are the most commonly used, however, lines with a range of ionization levels, like Mg\,II, C\,III] and C\,IV can also be used for AGNs at higher redshifts \citep{2016MNRAS.460..187M}. A few decades of intense RM campaigns have shown that photometric radius scales well with continuum luminosity, which allows us to measure the SMBH from a single-epoch optical spectrum \citep[][for a review]{2014SSRv..183..253P}.

Another single-epoch method that is recently proposed, uses the rotation of the polarization plane position angle across the broad emission line profile in order to trace the Keplerian motion and determine the SMBH mass \citep[][hereafter \citetalias{2015ApJ...800L..35A}]{2015ApJ...800L..35A}. It assumes that the BLR is flattened and the light is dominantly being scattered from the inner side of the dusty torus \citep[equatorial scattering,][]{2005MNRAS.359..846S}, resulting in the broad line polarization. This method additionally requires that the distance to the scattering region (SR) is known, whether using dust RM in the infrared \citep{2014ApJ...784L...4H,2014ApJ...788..159K} or measured directly using the infrared interferometry \citep{2011A&A...536A..78K}. In the latter case, \citetalias{2015ApJ...800L..35A} and the RM single-epoch method use different input observables, which makes it plausible to assume that these two methods are independent.

Detailed investigation of the \citetalias{2015ApJ...800L..35A} method by \citet{2018A&A...614A.120S,2019IJCAA...1...50S} have shown that it can be used when outflow/inflow velocity components are present, but low. Subsequently, \citet{2019MNRAS.482.4985A} have used the \citetalias{2015ApJ...800L..35A} for a sample of 30 type-1 AGNs. The same authors have also found viewing inclinations, maximal extents of the BLR and the index of the power-law emissivity, demonstrating that the \citetalias{2015ApJ...800L..35A} method can be used for calibration purpose since it is in good agreement with the $\ind{\mathcal{M}}{bh} - \sigma_*$ relation and the reverberation mapping. However, the \citetalias{2015ApJ...800L..35A} method has been applied so far only for nearby type-1 AGNs exploiting the polarization of H$\alpha$ spectral line, although it could also be applied to broad emission lines like Mg\,II, C\,III] and C\,IV. These lines are known for their slightly blueshifted peaks and very often asymmetric profiles with a larger excess in the blue part of line. Such line profiles are very often associated with the additional BLR complex motion as radial inflows and vertical outflows \citep{1982ApJ...263...79G,2005MNRAS.356.1029B}. The Mg\,II line is no exemption, and recently, \citet{2019MNRAS.484.3180P} have shown that a significant inflow/outflow velocity component of a few thousand of \SI{}{\kilo\meter\per\second} is present. Knowing that the polarization state is highly sensitive to geometry and kinematics \citep{2007A&A...465..129G}, the presence of high inflowing/outflowing components in the BLR should have a strong influence on the polarization of the Mg\,II line.

In order to probe the \citetalias{2015ApJ...800L..35A} for Mg\,II line, we model the equatorial scattering for H$\alpha$, H$\beta$ and Mg\,II lines, and discuss the general polarization signature. The paper is organized as followed: in Section \ref{s:model} we describe the model and we list all the parameters used. Our results are given in Section \ref{s:results}, together with the description of the observation procedure. Finally, we discuss the implications of our results and outline main conclusions.

\section{Model setup} \label{s:model}
We apply full 3D radiative transfer with polarization using a publicly available code \textsc{stokes} \citep{2007A&A...465..129G,2012A&A...548A.121M,2015A&A...577A..66M,2018A&A...615A.171M,2018A&A...611A..39R}. The program is suitable for dealing with complex geometry and kinematics of the model and treats multiple reprocessing events such as electron and dust scattering as well as dust absorption. The luminosity of the source is divided into a large number of photon packages (typically more than \SI{d7}{} per wavelength bin) and follow the input SED (power-law for the continuum or Lorentz-profile for the emitted broad line). For each emitted photon, the code follows it's path and computes Stokes parameters I, Q, U and V after each scattering. If there is no scattering region along the photon's path, the photon with it's polarization state is finally registered by one of the virtual detectors in the sky. The total (unpolarized) flux (TF), polarization degree ($p$) and the polarization position angle $\varphi$ are computed by summing Stokes parameters of all detected photons for each spectral bin. The code was originally developed for modeling optical and UV scattering induced continuum polarization in the radio-quiet AGNs, but it can be applied for studying polarization of many astrophysical phenomena \citep{2014sf2a.conf..103M}. The default output of the code $\varphi = \ang{90}$ corresponds to a polarization state where electrical field vector E is oscillating in the direction parallel to the axis of the symmetry of the system (\textit{z}-axis). This is the opposite to the convention used by \citet{2005MNRAS.359..846S}.
\subsection{Model parameters}
We approximate the accretion disk emission with a point-like continuum source emitting isotropic\footnote{Although the emission of a thin accretion disk is in the form $\cos\theta$, this would not affect the obtained profiles itself, however we could expect a significant decrease in polarized flux due to seed photons having direction preference towards pole on viewing angles.} unpolarized radiation for which spectral energy distribution (SED) is given by a power-law $F_C \propto \nu^{-\alpha}$. We set $\alpha = 2$ which corresponds to a flat spectrum when frequency is substituted with wavelength.
\begin{figure}
   \centering
   \includegraphics[width=\hsize]{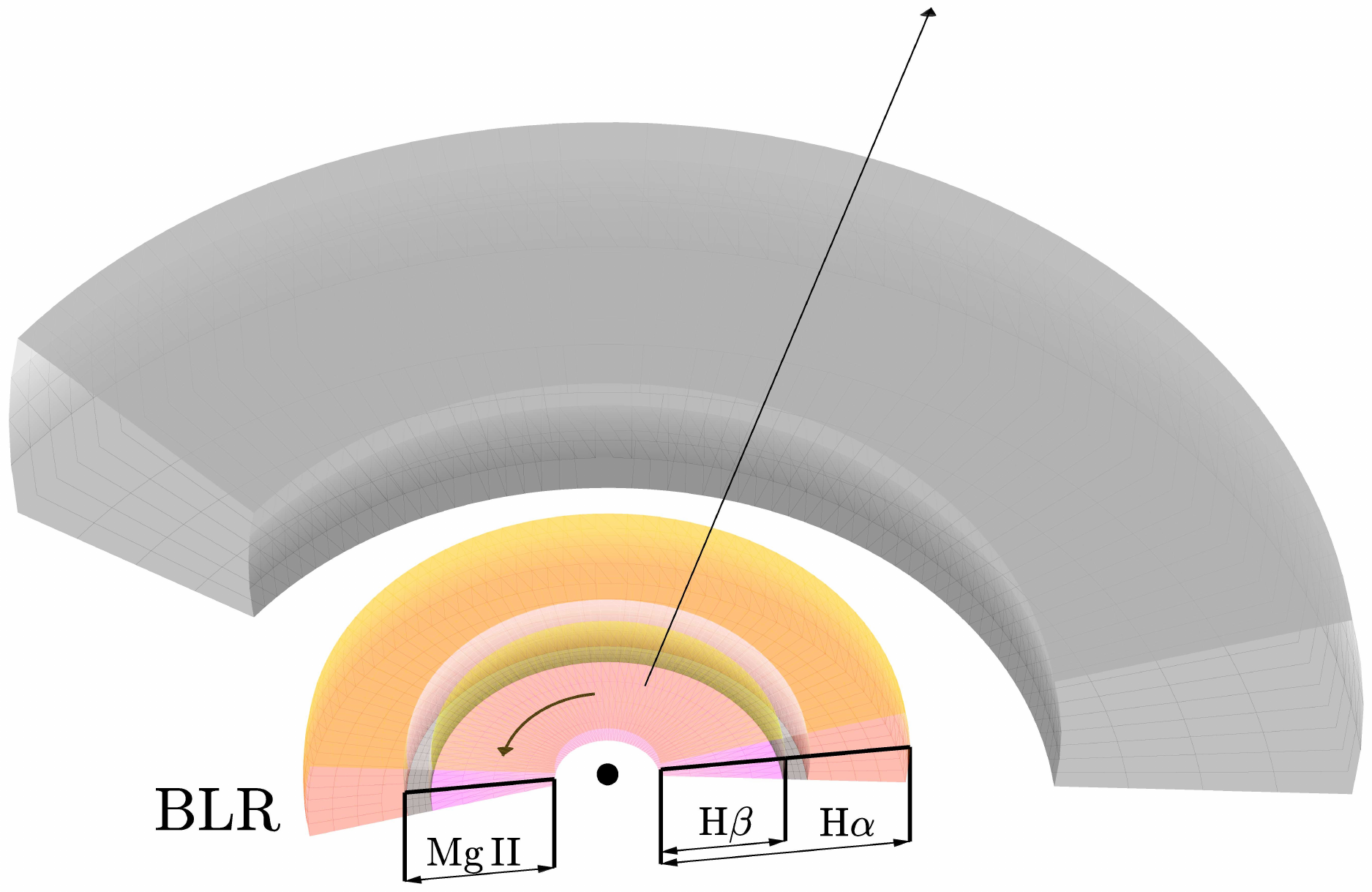}
      \caption{A 3D sketch showing the model geometry and kinematics of the three (H$\alpha$, H$\beta$ and Mg\,II) BLRs (orange) and the scattering region (grey). The size of each BLR is denoted with corresponding arrows and additional velocity component is accounted for the BLR of Mg\,II.}
   \label{f:model}
\end{figure}

The most convenient method for finding the size of the BLR is the reverberation mapping (RM) technique \citep{2005ApJ...629...61K,2006ApJ...651..775B,2013ApJ...767..149B}. \citet{2018A&A...614A.120S} tave compiled the RM measurement values found in literature for well known Type-1 AGNs and their luminosities at $\SI{5100}{\angstrom}$ ($\ind{L}{5100}$), and roughly estimated the BLR size (inner and outer radius) depending only on the mass of the SMBH. In this work, we set the SMBH mass to be $\ind{M}{bh} = \SI{d8}{\solarmass}$ and we adopt the same values for the corresponding H$\alpha$ BLR inner and outer radius (see Table \ref{t:mpar}). The corresponding BLR velocity is of the order of few thousands of \SI{}{\kilo\meter\per\second}. 

A systematic study by \citet{2019MNRAS.484.3180P} of 287 Type-1 AGNs with broad emission lines with redshift $0.407 < z < 0.643$ (in order to include both H$\beta$ and Mg\,II spectral lines), has shown that the Mg\,II BLR might be slightly larger than the H$\beta$ BLR since the FWHM of Mg\,II is slightly less than the FWHM of H$\beta$. We set the outer size of the Mg\,II BLR to be 10\% larger than the one for H$\beta$. The BLR was modeled as a distribution of gas in a disk-like flattened geometry with Keplerian motion with notable inflows and outflows present in the Mg\,II line. The complex structure of the BLR has been extensively studied via comparison of the broad line profiles between H$\beta$, Mg\,II and other spectral lines \citep[][and references therein]{2015ApJS..221...35K}. The RM measurements of optical Balmer lines for nearby Type-1 AGNs \citep{2010ApJ...716..993B} have shown that for most of the objects, the H$\alpha$ BLR is larger than the H$\beta$ BLR. From a much larger RM sample of Type-1 AGNs, the size of the Mg\,II BLR is consistently slightly larger than the size of the H$\beta$ BLR \citep{2016ApJ...818...30S}, which is in agreement with H$\beta$ being slightly more variable than Mg\,II line \citep{2015ApJ...811...42S}. For the sake of the model, in order to reduce the number of free parameters concerning the size of each BLR, we fix the size of the H$\beta$ BLR to be 50\% the size of the H$\alpha$ BLR and Mg\,II BLR to be 60\% the size of H$\alpha$ BLR (Fig.\ref{f:model}). The half opening angle for the BLR is \ang{30}, which correspond to the covering factor $\ind{CF}{BLR}= 0.5$. We assume that the BLR is transparent i.e.\,we neglect the line scattering by the BLR itself since the optical depth for Thomson scattering in our case is $\ind{\tau}{BLR} = 0.04\ind{R}{0.1\SI{}{\parsec}}$, where $\ind{R}{0.1\SI{}{\parsec}} = \ind{R}{BLR}/\SI{0.1}{\parsec}$ \citep{2018MNRAS.473L...1S}. For all three regions Keplerian motion is included. Only for the Mg\,II, constant \SI{6000}{\kilo\meter\per\second} inflow and outflow velocity component was added for the innermost one third of the region at an angle of \ang{60} with respect to the equatorial plane\footnote{Only the inflow/outflow velocity component was added while the same geometry of the Mg\,II region was kept.}.

In the work by \citet{2018A&A...614A.120S} it was found that the SR requires much higher covering factor and higher radial optical depth than the one used by \citet{2005MNRAS.359..846S} in order to produce the polarization signal typically observed in Type-1 AGNs. Assuming that equatorial scattering occurs only from the inner part of the torus, we adopt the same values for the SR radial thickness as given by \citet{2018A&A...614A.120S} with total radial optical depth equal to 1 for Thomson scattering. The half opening angle for the SR is \ang{35}, which corresponds to $\ind{CF}{SR} = 0.57$. The best SMBH mass estimates using polarization of broad emission lines are when the ratio between the SR inner radius and the BLR outer radius $\Rblr{SR}{in}/\Rblr{BLR}{out}$ is between 1.5 and 2.5. A value of \SI{1.72\pm0.48}{} for this ratio has been obtained by \citet{2019MNRAS.482.4985A}. Therefore we set the SR to be at twice the distance of the H$\alpha$ BLR when measured from the center. List of all model parameters is given in the Table \ref{t:mpar}. An illustration of the model geometry is shown in Fig.\ref{f:model}. We performed three separate simulations covering each of the Mg\,II, H$\beta$ and H$\alpha$ spectral domains.


\begin{table}
\caption{The inner and the outer radius of the BLRs for H$\alpha$, H$\beta$ and Mg\,II as well as for the SR. Spectral range and spectral resolution for each simulation around the central wavelengths.}
\centering
\begin{tabular}{ccccccc}
\toprule\toprule
& \multirow{2}{*}{Region}    & \multirow{2}{*}{$\Rblr{in}{}$} & \multirow{2}{*}{$\Rblr{out}{}$} & \multirow{2}{*}{$\ind{\lambda}{min}$} & \multirow{2}{*}{$\ind{\lambda}{max}$} & \multirow{2}{*}{spec.}\\
& & & & & & \raisebox{-0.3em}{res.}\\
& & ld & ld & $\SI{}{\angstrom}$ & $\SI{}{\angstrom}$ &\\
\midrule
     & H$\alpha$ & 36.94 & 58.93 & 6300 & 6826 & 300\\
 BLR & H$\beta$  & 36.94 & 47.91 & 4666 & 5055 & 300\\
     & Mg\,II    & 36.94 & 50.11 & 2688 & 2912 & 300\\
\midrule
 SR  &           & 117.87  & 201.22  \\
\bottomrule
\end{tabular}
\label{t:mpar}
\end{table}

\section{Results}\label{s:results}
In this section we compare polarization and line profiles for H$\alpha$, H$\beta$ and Mg\,II lines. The equatorial scattering dominates the systems with inclination range between \ang{20} and \ang{70}. We restrict viewing inclinations for Type-1 objects, which is in our case between \ang{20} and \ang{55}.

\begin{figure*}
  \centering
  \includegraphics[width=\hsize]{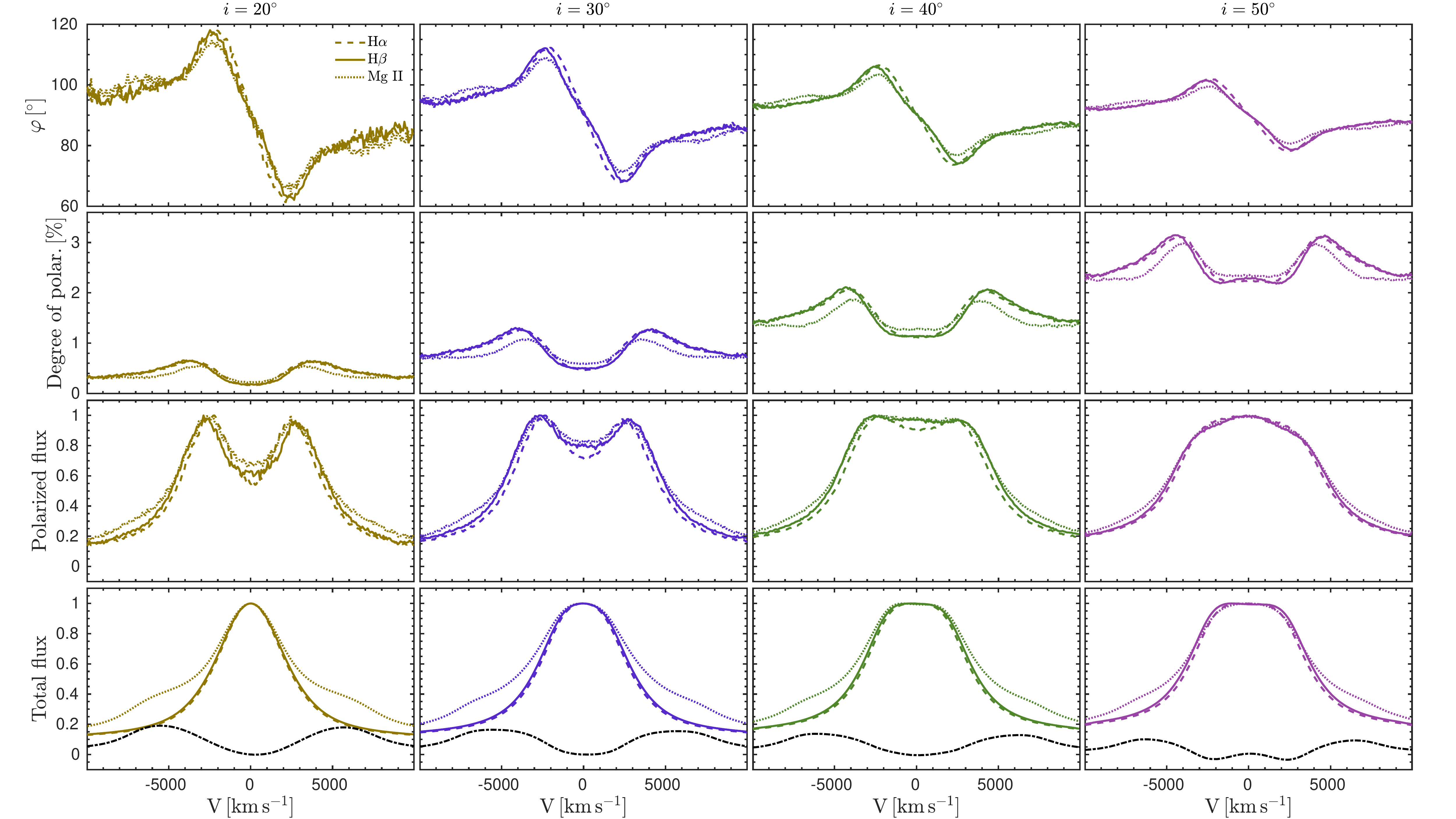}
  \caption{Polarization signature of each line for four viewing inclinations. Line styles corresponds to the following order: dash-dotted line is for H$\alpha$, solid line for H$\beta$ and dotted line for Mg\,II. The $\varphi$-profiles (top panels), degree of polarization (second from top), polarized flux (third from top) and total flux (bottom panels) are shown with the respect to velocity. Dashed black line (bottom panels) represent the difference between the Mg\,II and H$\beta$ unpolarized flux. Columns from the left to the right correspond to viewing inclination in ascending order, from near face-on towards intermediate inclinations.}
   \label{f:sve}%
\end{figure*}

In Fig.\,\ref{f:sve} (top panels), the profiles for $\varphi$ for each line and for four viewing inclinations. We can see that the $\varphi$-profiles for H$\alpha$ and H$\beta$ are nearly identical in the wings, while in the core, the position of the $\varphi$ amplitude (maximal offset from the continuum level which is $\ind{\varphi}{cont} = \ang{90}$) are for H$\alpha$ slightly shifted towards the core for roughly \SI{500}{\kilo\meter\per\second}. This is expected since the H$\alpha$ BLR is larger than the H$\beta$ BLR. The $\varphi$ amplitude for Mg\,II is around \ang{5} lower than the amplitudes for H$\alpha$ and H$\beta$. In the wings, the $\varphi$ amplitude for Mg\,II is showing a ``plateau'' rather than following profiles for pure Keplerian motion.

In Fig.\,\ref{f:sve} (top second panels) the results for simulated $p$ are shown. The double-peaked profile mentioned before is present for all three spectral lines. The $p$ profiles for H$\alpha$ and H$\beta$ are almost the same. The $p$ profile for Mg\,II shows lower polarization in the wings and slightly higher in the core than the $p$ for H$\alpha$ and H$\beta$. The $p$ maxima for Mg\,II are shifted towards blue for approximately \SI{1000}{\kilo\meter\per\second} with the respect to the maxima for the $p$ of H$\beta$ when viewed from the lowest viewing inclination (Fig.\,\ref{f:sve}, top second panels, first from the left). This shift of the maximum $p$ between Mg\,II and H$\beta$ (or H$\alpha$) is decreasing when the system is viewed from intermediate inclinations since the effects of the inflows and outflows are the greatest for the pole-on view.

In Fig.\,\ref{f:sve} (bottom second panels) the results for simulated PF are shown for all four viewing inclinations. Polarized lines look very similar for all three lines except that the polarized Mg\, II line is slightly stronger in the wings. In this case, the SR can fully resolve the Keplerian motion in the BLR, while the influence of the inflows and outflows present in the Mg\,II region are minor since the projection of the inflow or outflow velocity component in any direction towards the SR is much smaller in comparison with the Keplerian velocity. The polarized lines get broader when viewed from pole-on view towards the more inclined viewing angles and show a clear double-peaked profiles.

The results for unpolarized lines are shown in Fig.\,\ref{f:sve} (bottom panels). All profiles are single-peaked and broader when viewed from pole-on towards higher viewing inclinations. The profiles for H$\alpha$ and H$\beta$ lines are almost the same. The FWHM of H$\alpha$ line is less than the FWHM of H$\beta$ by \SI{500}{\kilo\meter\per\second}. This might be counter-intuitive since the H$\alpha$ BLR is twice the size of the H$\beta$ BLR. The reason is that for our model setup, the velocity difference between the outer parts of the H$\beta$ and H$\alpha$ BLRs is only \SI{300}{\kilo\meter\per\second} which combined with the inclination effects give slightly broader H$\beta$ than H$\alpha$ line. The effects of inflows and outflows present in the Mg\,II region is clearly visible in the strong wings of the Mg\,II line profile. Strong wings directly influence the $p$ profiles ($p = \mathrm{PF}/\mathrm{TF}$) by reducing net polarization in the Mg\,II line. The comparison between the H$\beta$ and Mg\,II lines is shown in Fig.\,\ref{f:sve} (bottom panels, dash-dotted line). It shows a symmetric double-peaked feature, very similar to the results by \citet{2019MNRAS.484.3180P} for the SDSS sample. We point out that the unpolarized lines are symmetric since the BLR is transparent in our model and we observe radiation from both sides of the equatorial plane instead of observing only the radiation coming from the side closer to the observer. Thus, both blue and red wings of the Mg\,II lines are prominent instead of having blue asymmetry that corresponds to a more realistic geometry.

\begin{figure*}
  \centering
  \includegraphics[width=\hsize]{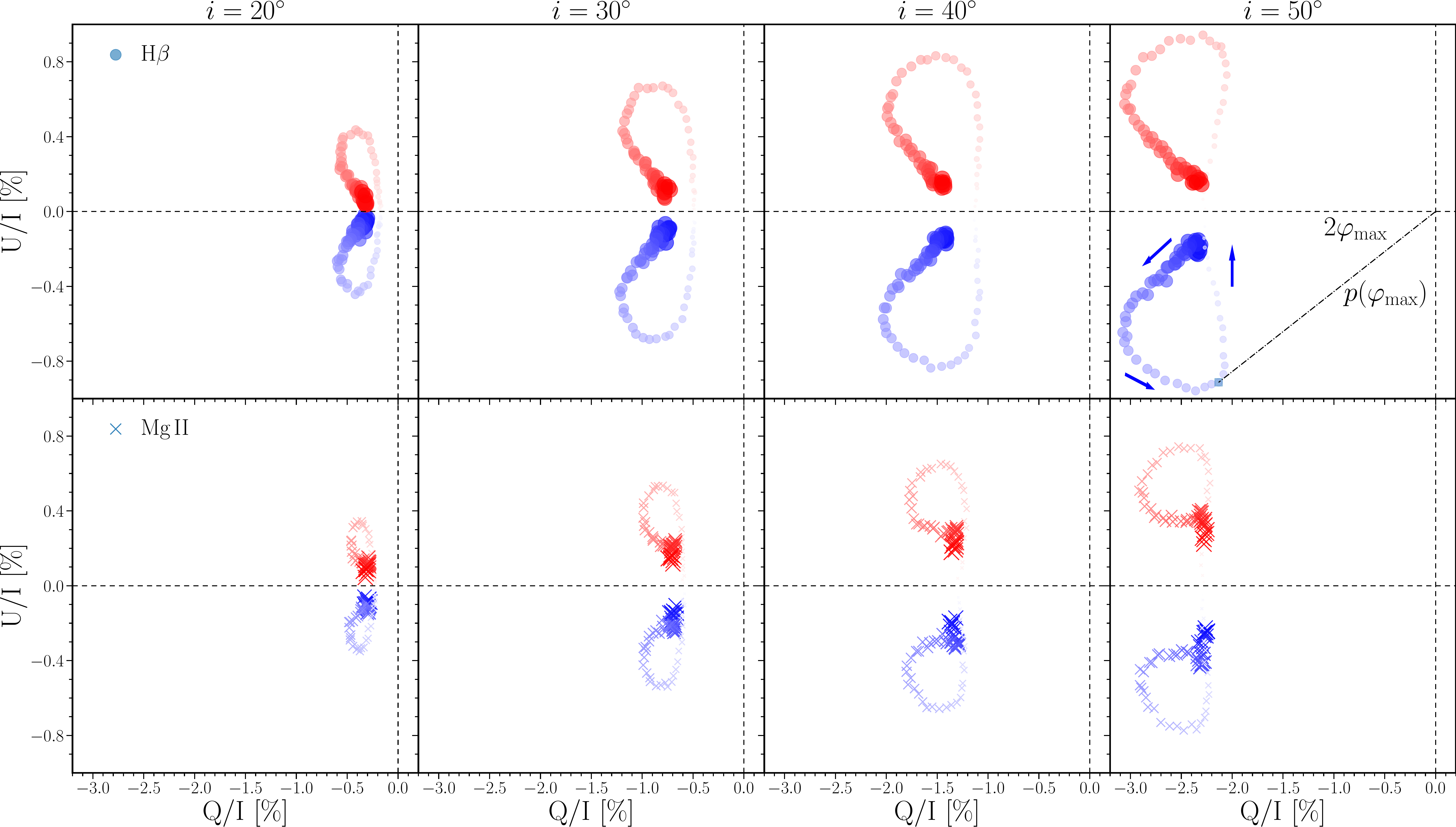}
  \caption{$QU$-plane for H$\beta$ (circles, upper panels) and Mg\,II (crosses, lower panels) normalized with $I$. Size and shade of symbols correspond to velocity in such way that greater size and darker shade correspond to higher velocities. Red denotes velocity greater than zero, while blue is the opposite. Dashed black lines are constant zero values of $Q$ and $U$. Panels from left to right are for four viewing inclinations. On the upper rightmost panel, blue arrows denote the direction of $Q$ and $U$ across the line profile. Blue square correspond to \ind{\varphi}{max} for H$\beta$ line. The angle between the dashed-dotted line and the $U = 0$ is $2\ind{\varphi}{max}$. The distance from the coordinate system origin represents $p$. Different symbols for H$\beta$ (circles) and Mg\,II (crosses) were used for contrast highlighting.}
   \label{f:QU}%
\end{figure*}

The $QU$-plane for H$\beta$ and Mg\,II line is shown in Fig.\,\ref{f:QU} for four viewing inclinations. In the same figure (upper rightmost panel), the evolution of the $Q$ and $U$ parameters along the line is indicated by blue arrows. The $U$ parameter starts around values close to zero and then it evolves giving rise to $\varphi$. When \ind{\varphi}{max} is reached, $U$ increases almost vertically and gets positive when line center is crossed. The opposite pattern is then followed in the red part of the line. In line wings, we can see that there is a clear distinction between the two groups of points for H$\beta$ and Mg\,II. The distance of each point from the center corresponds to $p$. Since $p$ in the wings is higher for H$\beta$ than for Mg\,II (Fig.\,\ref{f:sve}, top second panels), the $Q$ and $U$ parameters for H$\beta$ encompass the $Q$ and $U$ for Mg\,II in the $QU$-plane.

\subsection{Mass estimates}

The model predicts S-shaped profile of the polarization angle (Fig.\,\ref{f:sve}, top panels), which reflects Keplerian-like motion when equatorial scattering is a dominant scattering mechanism. Then, as it was shown in \citet{2014MNRAS.440..519A,2015ApJ...800L..35A}, velocity $V$ and polarization plane position angle $\varphi$ are connected by the following relation:
\begin{equation}
\log \left(\frac{V}{c} \right) = a - b \cdot \log(\tan[\Delta \varphi]),
\label{e:V_fi}
\end{equation}
where $c$ is the speed of light, $\Delta \varphi = \varphi - \langle\varphi\rangle$ is the difference between the polarization angle and its mean value and $a$ and $b$ are the coefficients of the linear approximation. The coefficient $b$ is equal to 0.5 as we assume the Keplerian-like motion. It is known that $a$ is connected with the BH mass \ind{\mathcal{M}}{bh} as:
\begin{equation}
a = 0.5 \log \left( \frac{G\ind{\mathcal{M}}{bh} \cos^2(\theta)}{c^2 \ind{R}{sc}} \right),
\end{equation}
\noindent where $G$ is the gravitational constant, \ind{R}{sc} is the distance from the central BH to the SR and $\theta$ is an angle between the BLR and the SR.

\begin{figure*}
  \centering
  \begin{tabular}{lc}
    \raisebox{0.20\hsize}{\Large{H$\alpha$}} & \includegraphics[width=0.72\hsize]{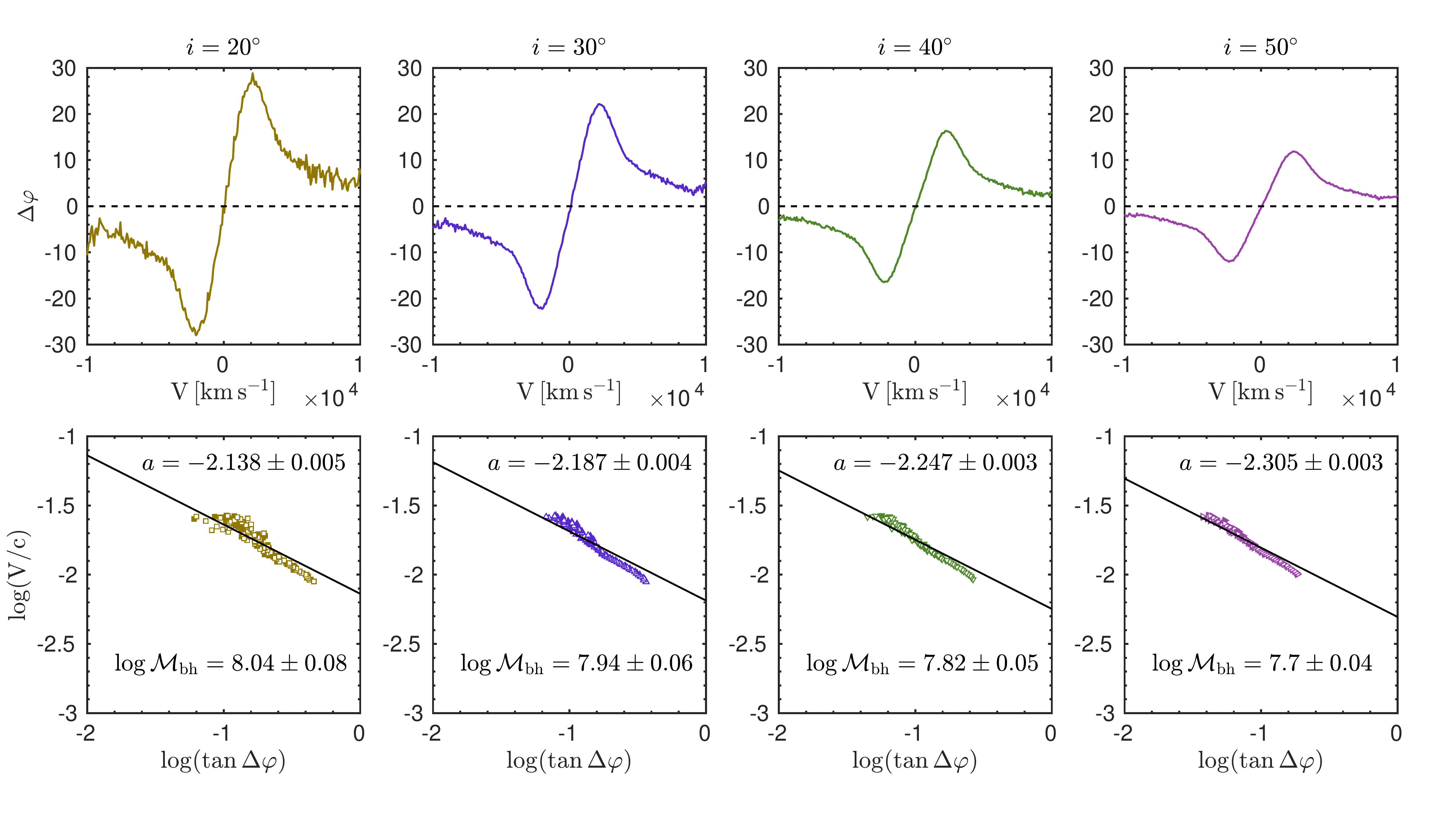} \\
    \midrule
    \raisebox{0.20\hsize}{\Large{H$\beta$}}  & \includegraphics[width=0.72\hsize]{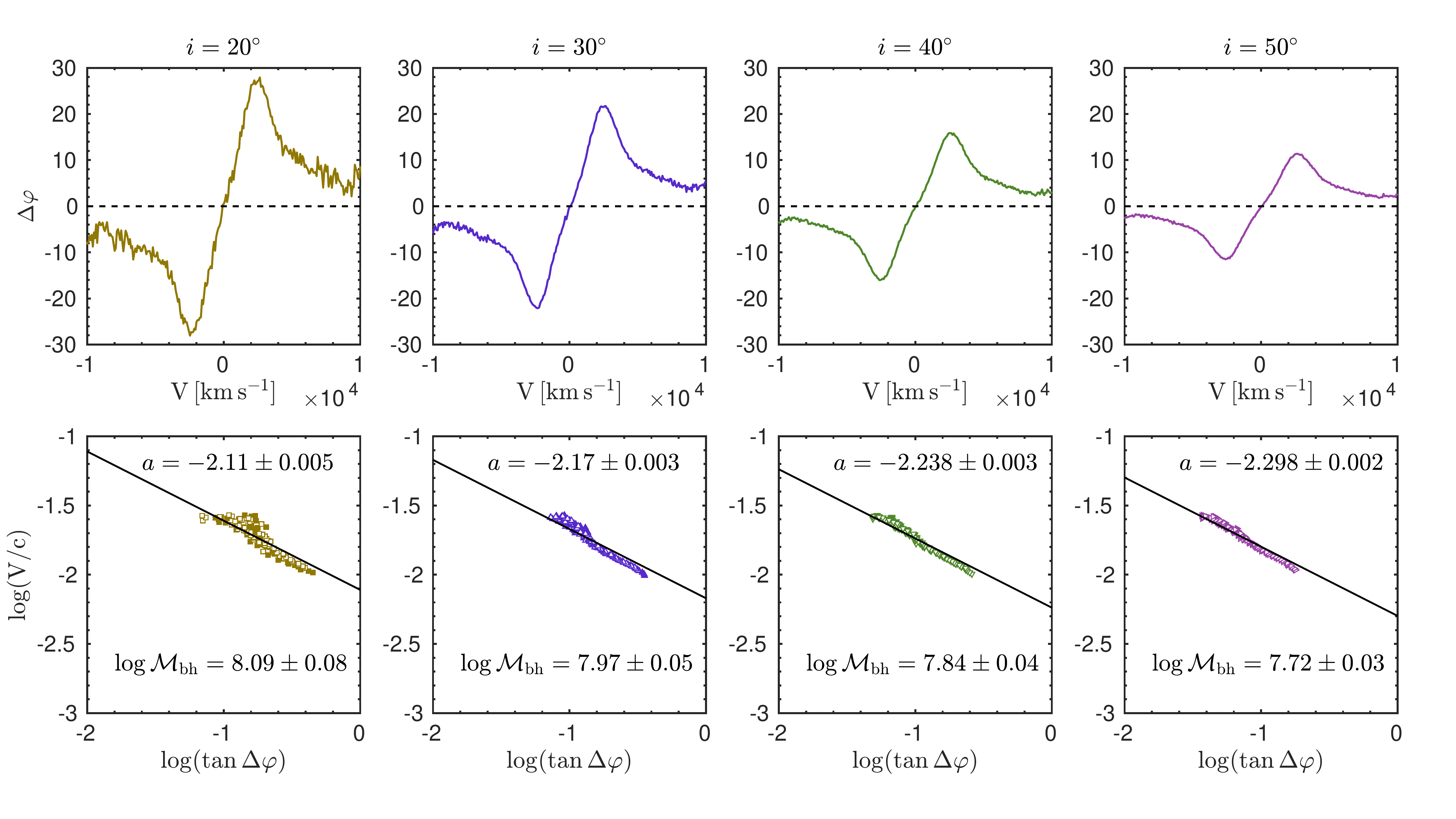} \\
    \midrule
    \raisebox{0.20\hsize}{\Large{Mg\,II}}    & \includegraphics[width=0.72\hsize]{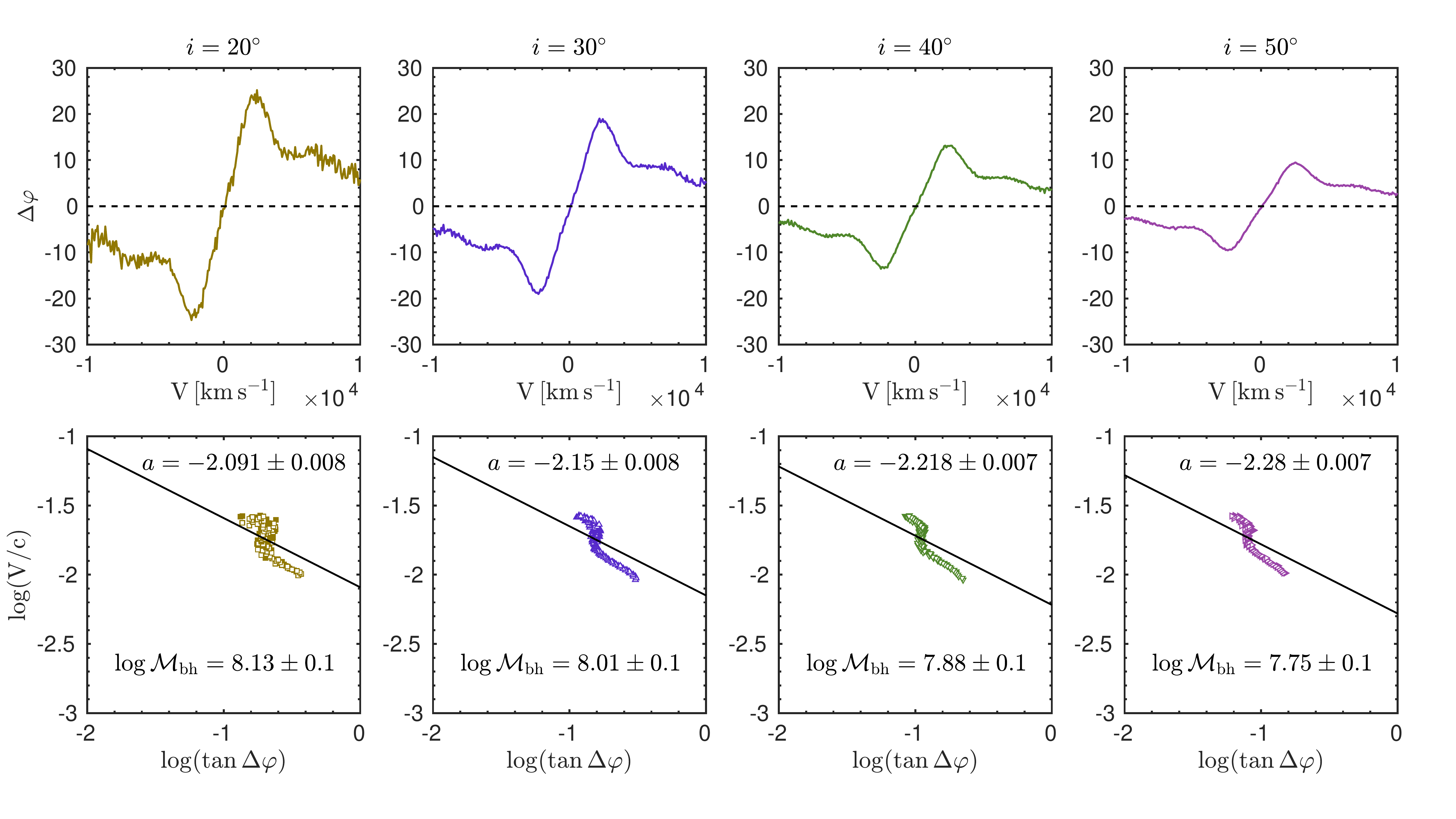}
  \end{tabular}
  \caption{SMBH mass estimates from the $\varphi$ of H$\alpha$ (top panels), H$\beta$ (middle panels) and Mg\,II (bottom panels). From left to right are viewing inclinations starting from \ang{20}, \ang{30}, \ang{40} and \ang{50}. For each line, panels are divided in two parts: upper part $\varphi$-profiles; lower part: $\log(V/c)-\log\tan\Delta\varphi$ linear fit. Empty and filled symbols in the lower part are for blue and red part of the line respectively.}
  \label{f:Mbh_model} 
\end{figure*}

In Fig.\,\ref{f:Mbh_model}, we show $\varphi$-profiles and linear fits using the equation \ref{e:V_fi} for all three spectral lines (H$\alpha$ top panels, H$\beta$ middle panels and Mg\,II bottom panels) and for four viewing inclinations (from left to right). We can see that for H$\alpha$ and H$\beta$ lines we obtain good linear fit, and mass estimates are close to the \SI{d8}{\solarmass} input mass. Mass estimates from H$\beta$ are systematically slightly higher than masses obtained from H$\alpha$ polarization angle profiles, owing to the H$\beta$ emission region having velocities that are up to \SI{500}{\kilo\meter\per\second} higher than the velocities of the H$\alpha$ emission region. In the case for Mg\,II line, the $\log (V/c) - \log\tan\Delta\varphi$ dependence significantly deviates from linear relation. We can see that linear relation (Keplerian motion) is valid only in the narrow velocity part between the peak and the plateau, which for our case corresponds to velocities between \SIrange{2500}{4500}{\kilo\meter\per\second} in both red and blue part of the line. The plateau covers the velocity range \SIrange{4500}{6500}{\kilo\meter\per\second} and a constant value of $\Delta\varphi$. This gives a vertical rise in the $\log (V/c) - \log\tan\Delta\varphi$, before $\Delta\varphi$ values finally drop to zero in the far wings. The Keplerian part almost matches the 1-$\sigma$ uncertainties when all points are used in the linear fit. If we perform linear fit only for these points, estimated SMBH masses are $\sim$\SI{35}{\percent} lower. From the observational point of view, the resolution is much worse and the data points are typically much more scattered around the straight line \citep{2019MNRAS.482.4985A} and observing $\varphi$-profiles similar to the modeled Mg\,II $\varphi$-profile would be difficult. Therefore, in a first approximation, we can perform a linear fit of the whole data set obtained from the observations of the polarized Mg\,II line, and assign additional \SI{35}{\percent} uncertainty to the estimated mass. That way the obtained the SMBH masses would still be of the same order with the masses estimated from the $\varphi$- profiles of H$\alpha$ or H$\beta$ lines where no or low velocity outflows are present. The exact values of parameter $a$ and SMBH masses obtained for linear fits using all points and for each viewing inclination are given in Table\,\ref{t:mase}.

\begin{table}
 \caption{SMBH masses estimated from H$\alpha$, H$\beta$ and Mg\,II lines for four viewing inclinations. Spectral line (Column 1), viewing inclinations (Column 2), parameter $a$ (Column 3), obtained masses given in \SI{}{\solarmass} (Column 4), estimated mass divided by input mass $\ind{\mathcal{M}}{input} = \SI{d8}{\solarmass}$ (Column 5).}
 \centering
 \begin{tabularx}{\columnwidth}{ccccc}
  \toprule\toprule
  \multicolumn{1}{c}{line} & $i(^\circ)$ & $a$ & $\log(\ind{\mathcal{M}}{bh}/\SI{}{\solarmass})$ & $\ind{\mathcal{M}}{bh}/\ind{\mathcal{M}}{input}$\\
  \midrule
  \multirow{3}{*}{H$\alpha$} & 20 & \SI{-2.138\pm0.005}{} & \SI{8.04\pm0.08}{} & 1.09 \\
                             & 30 & \SI{-2.187\pm0.004}{} & \SI{7.94\pm0.06}{} & 0.87 \\
                             & 40 & \SI{-2.247\pm0.003}{} & \SI{7.82\pm0.05}{} & 0.66 \\
                             & 50 & \SI{-2.305\pm0.003}{} & \SI{7.70\pm0.04}{} & 0.50 \\
  \midrule
  \multirow{3}{*}{H$\beta$}  & 20 & \SI{-2.110\pm0.005}{} & \SI{8.09\pm0.08}{} & 1.23 \\
                             & 30 & \SI{-2.170\pm0.003}{} & \SI{7.97\pm0.05}{} & 0.93 \\
                             & 40 & \SI{-2.238\pm0.003}{} & \SI{7.84\pm0.04}{} & 0.69 \\
                             & 50 & \SI{-2.298\pm0.002}{} & \SI{7.72\pm0.03}{} & 0.52 \\
  \midrule
  \multirow{3}{*}{Mg\,II}    & 20 & \SI{-2.091\pm0.008}{} & \SI{8.13\pm0.10}{} & 1.35 \\
                             & 30 & \SI{-2.150\pm0.008}{} & \SI{8.01\pm0.10}{} & 1.02 \\
                             & 40 & \SI{-2.218\pm0.007}{} & \SI{7.88\pm0.10}{} & 0.76 \\
                             & 50 & \SI{-2.280\pm0.007}{} & \SI{7.75\pm0.10}{} & 0.56 \\
  \bottomrule
 \end{tabularx}
 \label{t:mase}
\end{table}

\section{Discussion}
We investigated polarization effects in all three broad lines, focusing on the Mg\,II line and the application of the \citetalias{2015ApJ...800L..35A} method. The polarized lines have almost the same profiles and widths, for all three emission lines, even for such high inflows and outflows present in the Mg\,II BLR. The reason is that the emitted BLR radiation is seen by scatterers at close to edge-on viewing angles, and the projected vertical velocity component becomes low. The H$\beta$ and H$\alpha$ lines show almost identical $\varphi$, $p$, PF and TF profiles with differences in broadening effects of the order of \SI{500}{\kilo\meter\per\second}. SMBH mass estimates using H$\beta$ is $\sim$\SI{7}{\percent} higher than the one obtained using H$\alpha$ due to the smaller size of the H$\beta$ region. The Mg\,II emission line shows a plateau of constant $\varphi$ before dropping to the continuum value \ind{\varphi}{c} in the extreme line wings. In a first attempt, SMBH mass estimates from the Mg\,II emission line with extreme outflows would have additional $\sim$\SI{35}{\percent} error when compared with results obtained by using \citetalias{2015ApJ...800L..35A} method for H$\alpha$ and H$\beta$, which is still in agreement with previous results.

Single-epoch SMBH mass estimates using Mg\,II and C\,IV is of great importance for highly redshifted AGNs. Typically, SMBH mass using these lines is derived from the $\ind{L}{5100}-\ind{R}{BLR}$ relation for H$\beta$ line \citep{2006ApJ...641..689V,2009ApJ...707.1334W,2012MNRAS.427.3081T,2013A&A...555A..89M,2013ApJ...764..150M,2016MNRAS.460..187M,2019MNRAS.484.3180P}. If the emission of the Mg\,II line is dominated by the virialized component, we could expect a good agreement between the SMBHs obtained using the \citetalias{2015ApJ...800L..35A} and the single-epoch SMBH mass estimates using Mg\,II line. However, a considerable amount of objects show systematically blueshifted and asymmetric Mg\,II line profiles which is dominated by a non-virial kinematics \citep{2016MNRAS.460..187M} and for which the $\mathrm{FWHM} > \SI{6000}{\kilo\meter\per\second}$ \citep{2019MNRAS.484.3180P}. For these objects, we could expect much different geometry than the simple one we used.

When comparing the \citetalias{2015ApJ...800L..35A} method with the single-epoch SMBH mass estimates using FWHM, it is commonly assumed that the BLR gas is virialized in the vicinity of the black hole. This may not always be the case due to the uncertain gas distribution or the presence of the outflowing winds of various origin \citep{2013ApJ...763L..36L,2018NatAs...2...63M}. These effects can be observed in the polarized spectra, which is the advantage of the \citetalias{2015ApJ...800L..35A} method, however observational evidence still needs to be confirmed.

\citet{2020MNRAS.491....1L} have included large \SI{3000}{\kilo\meter\per\second} bulk outflows in the scattering region. They showed that such configuration greatly affects the observed $\varphi$-profiles which deviates from the one obtained for pure Keplerian motion. In our model, we didn't include complex motions of the SR since it is sufficiently far enough for outflowing velocities to be comparable with Keplerian velocity that is around \SI{2000}{\kilo\meter\per\second}. Low-magnitude inflows/outflows can be neglected \citep{2018A&A...614A.120S}.

\section{Conclusion}
We assumed equatorial scattering of the inner side of the dusty torus to be the main UV/optical broad line polarization mechanism. We used 3D Monte Carlo radiative transfer code \textsc{stokes} for accurate polarization treatment. We modeled equatorial scattering simultaneously for H$\alpha$, H$\beta$ and Mg\,II emission lines.

From the results obtained in this work we may conclude the following:
\begin{itemize}
 \item The presence of vertical inflows and outflows in the BLR that is much higher than the Keplerian velocity produces a plateau in the polarization plane position angle profiles.
 \item The application of the \citetalias{2015ApJ...800L..35A} method is valid as a rough first approximation even for the extreme outflows of the BLR.
 \item Error obtained this way is around $\sim$\SI{35}{\percent}.
\end{itemize}

We have paved the way for the use of the \citetalias{2015ApJ...800L..35A} method for highly ionized lines. For the future work, we plan to observe a few objects covering Mg\,II, C\,III], C\,IV and L$\alpha$ spectral range, and compare the SMBH mass estimates with other single-epoch methods in order to obtain more general results.

\section*{Acknowledgements}
We thank an anonymous referee for his remarks, comments and helpful suggestions that improved this paper. This work was supported by the Ministry of Education and Science (Republic of Serbia) through the project \textnumero451-03-68/2020/14/20002, Russian Foundation for Basic Research (RFBR) grant \textnumero15-02-02101, \textnumero14-22-03006. V.\,L.\,Afanasiev and E.\,S.\,Shablovinskaya were supported by the Russian Science Foundation (project \textnumero20-12-00030 ``Investigation of geometry and kinematics of ionized gas in active galactic nuclei by polarimetry methods''). Đ.\,Savić thanks the RFBR for the realization of the three months short term scientific visit at SAO funded by the grant \textnumero19-32-50009.

\bibliography{bibliography} 

\begin{thebibliography}{}
\makeatletter
\relax
\def\mn@urlcharsother{\let\do\@makeother \do\$\do\&\do\#\do\^\do\_\do\%\do\~}
\def\mn@doi{\begingroup\mn@urlcharsother \@ifnextchar [ {\mn@doi@}
  {\mn@doi@[]}}
\def\mn@doi@[#1]#2{\def\@tempa{#1}\ifx\@tempa\@empty \href
  {http://dx.doi.org/#2} {doi:#2}\else \href {http://dx.doi.org/#2} {#1}\fi
  \endgroup}
\def\mn@eprint#1#2{\mn@eprint@#1:#2::\@nil}
\def\mn@eprint@arXiv#1{\href {http://arxiv.org/abs/#1} {{\tt arXiv:#1}}}
\def\mn@eprint@dblp#1{\href {http://dblp.uni-trier.de/rec/bibtex/#1.xml}
  {dblp:#1}}
\def\mn@eprint@#1:#2:#3:#4\@nil{\def\@tempa {#1}\def\@tempb {#2}\def\@tempc
  {#3}\ifx \@tempc \@empty \let \@tempc \@tempb \let \@tempb \@tempa \fi \ifx
  \@tempb \@empty \def\@tempb {arXiv}\fi \@ifundefined
  {mn@eprint@\@tempb}{\@tempb:\@tempc}{\expandafter \expandafter \csname
  mn@eprint@\@tempb\endcsname \expandafter{\@tempc}}}

\bibitem[\protect\citeauthoryear{{Afanasiev} \& {Popovi{\'c}}}{{Afanasiev} \&
  {Popovi{\'c}}}{2015}]{2015ApJ...800L..35A}
{Afanasiev} V.~L.,  {Popovi{\'c}} L.~{\v C}.,  2015, \mn@doi [\apjl]
  {10.1088/2041-8205/800/2/L35}, \href
  {http://cdsads.u-strasbg.fr/abs/2015ApJ...800L..35A} {800, L35}

\bibitem[\protect\citeauthoryear{{Afanasiev}, {Popovi{\'c}}, {Shapovalova},
  {Borisov}  \& {Ili{\'c}}}{{Afanasiev} et~al.}{2014}]{2014MNRAS.440..519A}
{Afanasiev} V.~L.,  {Popovi{\'c}} L.~{\v C}.,  {Shapovalova} A.~I.,  {Borisov}
  N.~V.,   {Ili{\'c}} D.,  2014, \mn@doi [\mnras] {10.1093/mnras/stu231}, \href
  {http://adsabs.harvard.edu/abs/2014MNRAS.440..519A} {440, 519}

\bibitem[\protect\citeauthoryear{{Afanasiev}, {Popovi{\'c}}  \&
  {Shapovalova}}{{Afanasiev} et~al.}{2019}]{2019MNRAS.482.4985A}
{Afanasiev} V.~L.,  {Popovi{\'c}} L.~{\v C}.,   {Shapovalova} A.~I.,  2019,
  \mn@doi [\mnras] {10.1093/mnras/sty2995}, \href
  {http://adsabs.harvard.edu/abs/2019MNRAS.482.4985A} {482, 4985}

\bibitem[\protect\citeauthoryear{{Antonucci}}{{Antonucci}}{1993}]{1993ARA&A..31..473A}
{Antonucci} R.,  1993, \mn@doi [\araa] {10.1146/annurev.aa.31.090193.002353},
  \href {http://adsabs.harvard.edu/abs/1993ARA%26A..31..473A} {31, 473}

\bibitem[\protect\citeauthoryear{{Bahcall}, {Kozlovsky}  \&
  {Salpeter}}{{Bahcall} et~al.}{1972}]{1972ApJ...171..467B}
{Bahcall} J.~N.,  {Kozlovsky} B.-Z.,   {Salpeter} E.~E.,  1972, \mn@doi [\apj]
  {10.1086/151300}, \href
  {https://ui.adsabs.harvard.edu/abs/1972ApJ...171..467B} {171, 467}

\bibitem[\protect\citeauthoryear{{Barth} et~al.,}{{Barth}
  et~al.}{2013}]{2013ApJ...769..128B}
{Barth} A.~J.,  et~al., 2013, \mn@doi [\apj] {10.1088/0004-637X/769/2/128},
  \href {https://ui.adsabs.harvard.edu/abs/2013ApJ...769..128B} {769, 128}

\bibitem[\protect\citeauthoryear{{Barth} et~al.,}{{Barth}
  et~al.}{2015}]{2015ApJS..217...26B}
{Barth} A.~J.,  et~al., 2015, \mn@doi [\apjs] {10.1088/0067-0049/217/2/26},
  \href {https://ui.adsabs.harvard.edu/abs/2015ApJS..217...26B} {217, 26}

\bibitem[\protect\citeauthoryear{{Baskin} \& {Laor}}{{Baskin} \&
  {Laor}}{2005}]{2005MNRAS.356.1029B}
{Baskin} A.,  {Laor} A.,  2005, \mn@doi [\mnras]
  {10.1111/j.1365-2966.2004.08525.x}, \href
  {https://ui.adsabs.harvard.edu/abs/2005MNRAS.356.1029B} {356, 1029}

\bibitem[\protect\citeauthoryear{{Bentz} \& {Katz}}{{Bentz} \&
  {Katz}}{2015}]{2015PASP..127...67B}
{Bentz} M.~C.,  {Katz} S.,  2015, \mn@doi [\pasp] {10.1086/679601}, \href
  {http://adsabs.harvard.edu/abs/2015PASP..127...67B} {127, 67}

\bibitem[\protect\citeauthoryear{{Bentz} et~al.,}{{Bentz}
  et~al.}{2006}]{2006ApJ...651..775B}
{Bentz} M.~C.,  et~al., 2006, \mn@doi [\apj] {10.1086/507417}, \href
  {http://adsabs.harvard.edu/abs/2006ApJ...651..775B} {651, 775}

\bibitem[\protect\citeauthoryear{{Bentz} et~al.,}{{Bentz}
  et~al.}{2010}]{2010ApJ...716..993B}
{Bentz} M.~C.,  et~al., 2010, \mn@doi [\apj] {10.1088/0004-637X/716/2/993},
  \href {https://ui.adsabs.harvard.edu/abs/2010ApJ...716..993B} {716, 993}

\bibitem[\protect\citeauthoryear{{Bentz} et~al.,}{{Bentz}
  et~al.}{2013}]{2013ApJ...767..149B}
{Bentz} M.~C.,  et~al., 2013, \mn@doi [\apj] {10.1088/0004-637X/767/2/149},
  \href {http://adsabs.harvard.edu/abs/2013ApJ...767..149B} {767, 149}

\bibitem[\protect\citeauthoryear{{Blandford} \& {McKee}}{{Blandford} \&
  {McKee}}{1982}]{1982ApJ...255..419B}
{Blandford} R.~D.,  {McKee} C.~F.,  1982, \mn@doi [\apj] {10.1086/159843},
  \href {http://adsabs.harvard.edu/abs/1982ApJ...255..419B} {255, 419}

\bibitem[\protect\citeauthoryear{{Clavel} et~al.,}{{Clavel}
  et~al.}{1991}]{1991ApJ...366...64C}
{Clavel} J.,  et~al., 1991, \mn@doi [\apj] {10.1086/169540}, \href
  {https://ui.adsabs.harvard.edu/abs/1991ApJ...366...64C} {366, 64}

\bibitem[\protect\citeauthoryear{{Du} \& {Wang}}{{Du} \&
  {Wang}}{2019}]{2019ApJ...886...42D}
{Du} P.,  {Wang} J.-M.,  2019, \mn@doi [\apj] {10.3847/1538-4357/ab4908}, \href
  {https://ui.adsabs.harvard.edu/abs/2019ApJ...886...42D} {886, 42}

\bibitem[\protect\citeauthoryear{{Du} et~al.,}{{Du}
  et~al.}{2014}]{2014ApJ...782...45D}
{Du} P.,  et~al., 2014, \mn@doi [\apj] {10.1088/0004-637X/782/1/45}, \href
  {https://ui.adsabs.harvard.edu/abs/2014ApJ...782...45D} {782, 45}

\bibitem[\protect\citeauthoryear{{Du} et~al.,}{{Du}
  et~al.}{2015}]{2015ApJ...806...22D}
{Du} P.,  et~al., 2015, \mn@doi [\apj] {10.1088/0004-637X/806/1/22}, \href
  {https://ui.adsabs.harvard.edu/abs/2015ApJ...806...22D} {806, 22}

\bibitem[\protect\citeauthoryear{{Du} et~al.,}{{Du}
  et~al.}{2018}]{2018ApJ...856....6D}
{Du} P.,  et~al., 2018, \mn@doi [\apj] {10.3847/1538-4357/aaae6b}, \href
  {http://adsabs.harvard.edu/abs/2018ApJ...856....6D} {856, 6}

\bibitem[\protect\citeauthoryear{{Fabian}}{{Fabian}}{2012}]{2012ARA&A..50..455F}
{Fabian} A.~C.,  2012, \mn@doi [\araa] {10.1146/annurev-astro-081811-125521},
  \href {http://adsabs.harvard.edu/abs/2012ARA%26A..50..455F} {50, 455}

\bibitem[\protect\citeauthoryear{{Gaskell}}{{Gaskell}}{1982}]{1982ApJ...263...79G}
{Gaskell} C.~M.,  1982, \mn@doi [\apj] {10.1086/160481}, \href
  {http://adsabs.harvard.edu/abs/1982ApJ...263...79G} {263, 79}

\bibitem[\protect\citeauthoryear{{Goosmann} \& {Gaskell}}{{Goosmann} \&
  {Gaskell}}{2007}]{2007A&A...465..129G}
{Goosmann} R.~W.,  {Gaskell} C.~M.,  2007, \mn@doi [\aap]
  {10.1051/0004-6361:20053555}, \href
  {http://adsabs.harvard.edu/abs/2007A%26A...465..129G} {465, 129}

\bibitem[\protect\citeauthoryear{{Grier} et~al.,}{{Grier}
  et~al.}{2017}]{2017ApJ...851...21G}
{Grier} C.~J.,  et~al., 2017, \mn@doi [\apj] {10.3847/1538-4357/aa98dc}, \href
  {https://ui.adsabs.harvard.edu/abs/2017ApJ...851...21G} {851, 21}

\bibitem[\protect\citeauthoryear{{Grier} et~al.,}{{Grier}
  et~al.}{2019}]{2019ApJ...887...38G}
{Grier} C.~J.,  et~al., 2019, \mn@doi [\apj] {10.3847/1538-4357/ab4ea5}, \href
  {https://ui.adsabs.harvard.edu/abs/2019ApJ...887...38G} {887, 38}

\bibitem[\protect\citeauthoryear{{Heckman} \& {Kauffmann}}{{Heckman} \&
  {Kauffmann}}{2011}]{2011Sci...333..182H}
{Heckman} T.~M.,  {Kauffmann} G.,  2011, \mn@doi [Science]
  {10.1126/science.1200504}, \href
  {https://ui.adsabs.harvard.edu/abs/2011Sci...333..182H} {333, 182}

\bibitem[\protect\citeauthoryear{{H{\"o}nig}}{{H{\"o}nig}}{2014}]{2014ApJ...784L...4H}
{H{\"o}nig} S.~F.,  2014, \mn@doi [\apjl] {10.1088/2041-8205/784/1/L4}, \href
  {https://ui.adsabs.harvard.edu/abs/2014ApJ...784L...4H} {784, L4}

\bibitem[\protect\citeauthoryear{{Ili{\'c}} et~al.,}{{Ili{\'c}}
  et~al.}{2017}]{2017FrASS...4...12I}
{Ili{\'c}} D.,  et~al., 2017, \mn@doi [Frontiers in Astronomy and Space
  Sciences] {10.3389/fspas.2017.00012}, \href
  {https://ui.adsabs.harvard.edu/abs/2017FrASS...4...12I} {4, 12}

\bibitem[\protect\citeauthoryear{{Kaspi}, {Smith}, {Netzer}, {Maoz}, {Jannuzi}
  \& {Giveon}}{{Kaspi} et~al.}{2000}]{2000ApJ...533..631K}
{Kaspi} S.,  {Smith} P.~S.,  {Netzer} H.,  {Maoz} D.,  {Jannuzi} B.~T.,
  {Giveon} U.,  2000, \mn@doi [\apj] {10.1086/308704}, \href
  {http://adsabs.harvard.edu/abs/2000ApJ...533..631K} {533, 631}

\bibitem[\protect\citeauthoryear{{Kaspi}, {Maoz}, {Netzer}, {Peterson},
  {Vestergaard}  \& {Jannuzi}}{{Kaspi} et~al.}{2005}]{2005ApJ...629...61K}
{Kaspi} S.,  {Maoz} D.,  {Netzer} H.,  {Peterson} B.~M.,  {Vestergaard} M.,
  {Jannuzi} B.~T.,  2005, \mn@doi [\apj] {10.1086/431275}, \href
  {http://adsabs.harvard.edu/abs/2005ApJ...629...61K} {629, 61}

\bibitem[\protect\citeauthoryear{{Kaspi}, {Brandt}, {Maoz}, {Netzer},
  {Schneider}  \& {Shemmer}}{{Kaspi} et~al.}{2007}]{2007ApJ...659..997K}
{Kaspi} S.,  {Brandt} W.~N.,  {Maoz} D.,  {Netzer} H.,  {Schneider} D.~P.,
  {Shemmer} O.,  2007, \mn@doi [\apj] {10.1086/512094}, \href
  {https://ui.adsabs.harvard.edu/abs/2007ApJ...659..997K} {659, 997}

\bibitem[\protect\citeauthoryear{{Kishimoto}, {H{\"o}nig}, {Antonucci},
  {Millour}, {Tristram}  \& {Weigelt}}{{Kishimoto}
  et~al.}{2011}]{2011A&A...536A..78K}
{Kishimoto} M.,  {H{\"o}nig} S.~F.,  {Antonucci} R.,  {Millour} F.,  {Tristram}
  K.~R.~W.,   {Weigelt} G.,  2011, \mn@doi [\aap]
  {10.1051/0004-6361/201117367}, \href
  {http://adsabs.harvard.edu/abs/2011A%26A...536A..78K} {536, A78}

\bibitem[\protect\citeauthoryear{{Kormendy} \& {Ho}}{{Kormendy} \&
  {Ho}}{2013}]{2013ARA&A..51..511K}
{Kormendy} J.,  {Ho} L.~C.,  2013, \mn@doi [\araa]
  {10.1146/annurev-astro-082708-101811}, \href
  {http://adsabs.harvard.edu/abs/2013ARA%26A..51..511K} {51, 511}

\bibitem[\protect\citeauthoryear{{Kormendy} \& {Richstone}}{{Kormendy} \&
  {Richstone}}{1995}]{1995ARA&A..33..581K}
{Kormendy} J.,  {Richstone} D.,  1995, \mn@doi [\araa]
  {10.1146/annurev.aa.33.090195.003053}, \href
  {http://adsabs.harvard.edu/abs/1995ARA%26A..33..581K} {33, 581}

\bibitem[\protect\citeauthoryear{{Koshida} et~al.,}{{Koshida}
  et~al.}{2014}]{2014ApJ...788..159K}
{Koshida} S.,  et~al., 2014, \mn@doi [\apj] {10.1088/0004-637X/788/2/159},
  \href {http://adsabs.harvard.edu/abs/2014ApJ...788..159K} {788, 159}

\bibitem[\protect\citeauthoryear{{Kova{\v{c}}evi{\'c}-Doj{\v{c}}inovi{\'c}} \&
  {Popovi{\'c}}}{{Kova{\v{c}}evi{\'c}-Doj{\v{c}}inovi{\'c}} \&
  {Popovi{\'c}}}{2015}]{2015ApJS..221...35K}
{Kova{\v{c}}evi{\'c}-Doj{\v{c}}inovi{\'c}} J.,  {Popovi{\'c}} L.~{\v{C}}.,
  2015, \mn@doi [\apjs] {10.1088/0067-0049/221/2/35}, \href
  {https://ui.adsabs.harvard.edu/abs/2015ApJS..221...35K} {221, 35}

\bibitem[\protect\citeauthoryear{{Le{\'o}n-Tavares} et~al.,}{{Le{\'o}n-Tavares}
  et~al.}{2013}]{2013ApJ...763L..36L}
{Le{\'o}n-Tavares} J.,  et~al., 2013, \mn@doi [\apjl]
  {10.1088/2041-8205/763/2/L36}, \href
  {https://ui.adsabs.harvard.edu/abs/2013ApJ...763L..36L} {763, L36}

\bibitem[\protect\citeauthoryear{{Lira}, {Goosmann}, {Kishimoto}  \&
  {Cartier}}{{Lira} et~al.}{2020}]{2020MNRAS.491....1L}
{Lira} P.,  {Goosmann} R.~W.,  {Kishimoto} M.,   {Cartier} R.,  2020, \mn@doi
  [\mnras] {10.1093/mnras/stz2774}, \href
  {https://ui.adsabs.harvard.edu/abs/2020MNRAS.491....1L} {491, 1}

\bibitem[\protect\citeauthoryear{{Lynden-Bell}}{{Lynden-Bell}}{1969}]{1969Natur.223..690L}
{Lynden-Bell} D.,  1969, \mn@doi [\nat] {10.1038/223690a0}, \href
  {http://adsabs.harvard.edu/abs/1969Natur.223..690L} {223, 690}

\bibitem[\protect\citeauthoryear{{Marin}}{{Marin}}{2018}]{2018A&A...615A.171M}
{Marin} F.,  2018, \mn@doi [\aap] {10.1051/0004-6361/201833225}, \href
  {https://ui.adsabs.harvard.edu/abs/2018A&A...615A.171M} {615, A171}

\bibitem[\protect\citeauthoryear{{Marin} \& {Goosmann}}{{Marin} \&
  {Goosmann}}{2014}]{2014sf2a.conf..103M}
{Marin} F.,  {Goosmann} R.~W.,  2014, in {Ballet} J.,  {Martins} F.,
  {Bournaud} F.,  {Monier} R.,   {Reyl{\'e}} C.,  eds, SF2A-2014: Proceedings
  of the Annual meeting of the French Society of Astronomy and Astrophysics. pp
  103--108 (\mn@eprint {arXiv} {1409.7278})

\bibitem[\protect\citeauthoryear{{Marin}, {Goosmann}, {Gaskell}, {Porquet}  \&
  {Dov{\v c}iak}}{{Marin} et~al.}{2012}]{2012A&A...548A.121M}
{Marin} F.,  {Goosmann} R.~W.,  {Gaskell} C.~M.,  {Porquet} D.,   {Dov{\v
  c}iak} M.,  2012, \mn@doi [\aap] {10.1051/0004-6361/201219751}, \href
  {http://adsabs.harvard.edu/abs/2012A%26A...548A.121M} {548, A121}

\bibitem[\protect\citeauthoryear{{Marin}, {Goosmann}  \& {Gaskell}}{{Marin}
  et~al.}{2015}]{2015A&A...577A..66M}
{Marin} F.,  {Goosmann} R.~W.,   {Gaskell} C.~M.,  2015, \mn@doi [\aap]
  {10.1051/0004-6361/201525628}, \href
  {http://adsabs.harvard.edu/abs/2015A%26A...577A..66M} {577, A66}

\bibitem[\protect\citeauthoryear{{Marziani}, {Sulentic}, {Plauchu-Frayn}  \&
  {del Olmo}}{{Marziani} et~al.}{2013a}]{2013A&A...555A..89M}
{Marziani} P.,  {Sulentic} J.~W.,  {Plauchu-Frayn} I.,   {del Olmo} A.,  2013a,
  \mn@doi [\aap] {10.1051/0004-6361/201321374}, \href
  {https://ui.adsabs.harvard.edu/abs/2013A&A...555A..89M} {555, A89}

\bibitem[\protect\citeauthoryear{{Marziani}, {Sulentic}, {Plauchu-Frayn}  \&
  {del Olmo}}{{Marziani} et~al.}{2013b}]{2013ApJ...764..150M}
{Marziani} P.,  {Sulentic} J.~W.,  {Plauchu-Frayn} I.,   {del Olmo} A.,  2013b,
  \mn@doi [\apj] {10.1088/0004-637X/764/2/150}, \href
  {https://ui.adsabs.harvard.edu/abs/2013ApJ...764..150M} {764, 150}

\bibitem[\protect\citeauthoryear{{Mej{\'\i}a-Restrepo}, {Trakhtenbrot}, {Lira},
  {Netzer}  \& {Capellupo}}{{Mej{\'\i}a-Restrepo}
  et~al.}{2016}]{2016MNRAS.460..187M}
{Mej{\'\i}a-Restrepo} J.~E.,  {Trakhtenbrot} B.,  {Lira} P.,  {Netzer} H.,
  {Capellupo} D.~M.,  2016, \mn@doi [\mnras] {10.1093/mnras/stw568}, \href
  {https://ui.adsabs.harvard.edu/abs/2016MNRAS.460..187M} {460, 187}

\bibitem[\protect\citeauthoryear{{Mej{\'\i}a-Restrepo}, {Lira}, {Netzer},
  {Trakhtenbrot}  \& {Capellupo}}{{Mej{\'\i}a-Restrepo}
  et~al.}{2018}]{2018NatAs...2...63M}
{Mej{\'\i}a-Restrepo} J.~E.,  {Lira} P.,  {Netzer} H.,  {Trakhtenbrot} B.,
  {Capellupo} D.~M.,  2018, \mn@doi [Nature Astronomy]
  {10.1038/s41550-017-0305-z}, \href
  {https://ui.adsabs.harvard.edu/abs/2018NatAs...2...63M} {2, 63}

\bibitem[\protect\citeauthoryear{{Netzer}}{{Netzer}}{2013}]{2013peag.book.....N}
{Netzer} H.,  2013, {The Physics and Evolution of Active Galactic Nuclei}.
Cambridge University Press

\bibitem[\protect\citeauthoryear{{Netzer}}{{Netzer}}{2015}]{2015ARA&A..53..365N}
{Netzer} H.,  2015, \mn@doi [\araa] {10.1146/annurev-astro-082214-122302},
  \href {http://adsabs.harvard.edu/abs/2015ARA%26A..53..365N} {53, 365}

\bibitem[\protect\citeauthoryear{{Peterson}}{{Peterson}}{1993}]{1993PASP..105..247P}
{Peterson} B.~M.,  1993, \mn@doi [\pasp] {10.1086/133140}, \href
  {https://ui.adsabs.harvard.edu/abs/1993PASP..105..247P} {105, 247}

\bibitem[\protect\citeauthoryear{{Peterson}}{{Peterson}}{2014}]{2014SSRv..183..253P}
{Peterson} B.~M.,  2014, \mn@doi [\ssr] {10.1007/s11214-013-9987-4}, \href
  {http://adsabs.harvard.edu/abs/2014SSRv..183..253P} {183, 253}

\bibitem[\protect\citeauthoryear{{Peterson} et~al.,}{{Peterson}
  et~al.}{2004}]{2004ApJ...613..682P}
{Peterson} B.~M.,  et~al., 2004, \mn@doi [\apj] {10.1086/423269}, \href
  {http://adsabs.harvard.edu/abs/2004ApJ...613..682P} {613, 682}

\bibitem[\protect\citeauthoryear{{Popovi{\'c}}, {Kova{\v c}evi{\'c}-Doj{\v
  c}inovi{\'c}}  \& {Mar{\v c}eta-Mandi{\'c}}}{{Popovi{\'c}}
  et~al.}{2019}]{2019MNRAS.484.3180P}
{Popovi{\'c}} L.~{\v C}.,  {Kova{\v c}evi{\'c}-Doj{\v c}inovi{\'c}} J.,
  {Mar{\v c}eta-Mandi{\'c}} S.,  2019, \mn@doi [\mnras] {10.1093/mnras/stz157},
  \href {https://ui.adsabs.harvard.edu/abs/2019MNRAS.484.3180P} {484, 3180}

\bibitem[\protect\citeauthoryear{{Rojas Lobos}, {Goosmann}, {Marin}  \&
  {Savi{\'c}}}{{Rojas Lobos} et~al.}{2018}]{2018A&A...611A..39R}
{Rojas Lobos} P.~A.,  {Goosmann} R.~W.,  {Marin} F.,   {Savi{\'c}} D.,  2018,
  \mn@doi [\aap] {10.1051/0004-6361/201731331}, \href
  {https://ui.adsabs.harvard.edu/abs/2018A&A...611A..39R} {611, A39}

\bibitem[\protect\citeauthoryear{{Salpeter}}{{Salpeter}}{1964}]{1964ApJ...140..796S}
{Salpeter} E.~E.,  1964, \mn@doi [\apj] {10.1086/147973}, \href
  {http://adsabs.harvard.edu/abs/1964ApJ...140..796S} {140, 796}

\bibitem[\protect\citeauthoryear{{Savi{\'c}}}{{Savi{\'c}}}{2019}]{2019IJCAA...1...50S}
{Savi{\'c}} D.,  2019, \mn@doi [International Journal of Cosmology]
  {10.18689/ijcaa-1000113}, \href
  {https://ui.adsabs.harvard.edu/abs/2019IJCAA...1...50S} {1, 50}

\bibitem[\protect\citeauthoryear{{Savi{\'c}}, {Goosmann}, {Popovi{\'c}},
  {Marin}  \& {Afanasiev}}{{Savi{\'c}} et~al.}{2018}]{2018A&A...614A.120S}
{Savi{\'c}} D.,  {Goosmann} R.,  {Popovi{\'c}} L.~{\v{C}}.,  {Marin} F.,
  {Afanasiev} V.~L.,  2018, \mn@doi [\aap] {10.1051/0004-6361/201732220}, \href
  {https://ui.adsabs.harvard.edu/abs/2018A&A...614A.120S} {614, A120}

\bibitem[\protect\citeauthoryear{{Shapovalova} et~al.,}{{Shapovalova}
  et~al.}{2009}]{2009NewAR..53..191S}
{Shapovalova} A.~I.,  et~al., 2009, \mn@doi [\nar]
  {10.1016/j.newar.2009.08.004}, \href
  {https://ui.adsabs.harvard.edu/abs/2009NewAR..53..191S} {53, 191}

\bibitem[\protect\citeauthoryear{{Shen} et~al.,}{{Shen}
  et~al.}{2016}]{2016ApJ...818...30S}
{Shen} Y.,  et~al., 2016, \mn@doi [\apj] {10.3847/0004-637X/818/1/30}, \href
  {https://ui.adsabs.harvard.edu/abs/2016ApJ...818...30S} {818, 30}

\bibitem[\protect\citeauthoryear{{Smith}, {Robinson}, {Young}, {Axon}  \&
  {Corbett}}{{Smith} et~al.}{2005}]{2005MNRAS.359..846S}
{Smith} J.~E.,  {Robinson} A.,  {Young} S.,  {Axon} D.~J.,   {Corbett} E.~A.,
  2005, \mn@doi [\mnras] {10.1111/j.1365-2966.2005.08895.x}, \href
  {http://adsabs.harvard.edu/abs/2005MNRAS.359..846S} {359, 846}

\bibitem[\protect\citeauthoryear{{Songsheng} \& {Wang}}{{Songsheng} \&
  {Wang}}{2018}]{2018MNRAS.473L...1S}
{Songsheng} Y.-Y.,  {Wang} J.-M.,  2018, \mn@doi [\mnras]
  {10.1093/mnrasl/slx154}, \href
  {http://adsabs.harvard.edu/abs/2018MNRAS.473L...1S} {473, L1}

\bibitem[\protect\citeauthoryear{{Sun} et~al.,}{{Sun}
  et~al.}{2015}]{2015ApJ...811...42S}
{Sun} M.,  et~al., 2015, \mn@doi [\apj] {10.1088/0004-637X/811/1/42}, \href
  {https://ui.adsabs.harvard.edu/abs/2015ApJ...811...42S} {811, 42}

\bibitem[\protect\citeauthoryear{{Trakhtenbrot} \& {Netzer}}{{Trakhtenbrot} \&
  {Netzer}}{2012}]{2012MNRAS.427.3081T}
{Trakhtenbrot} B.,  {Netzer} H.,  2012, \mn@doi [\mnras]
  {10.1111/j.1365-2966.2012.22056.x}, \href
  {https://ui.adsabs.harvard.edu/abs/2012MNRAS.427.3081T} {427, 3081}

\bibitem[\protect\citeauthoryear{{Vestergaard} \& {Peterson}}{{Vestergaard} \&
  {Peterson}}{2006}]{2006ApJ...641..689V}
{Vestergaard} M.,  {Peterson} B.~M.,  2006, \mn@doi [\apj] {10.1086/500572},
  \href {https://ui.adsabs.harvard.edu/abs/2006ApJ...641..689V} {641, 689}

\bibitem[\protect\citeauthoryear{{Wang} et~al.,}{{Wang}
  et~al.}{2009}]{2009ApJ...707.1334W}
{Wang} J.-G.,  et~al., 2009, \mn@doi [\apj] {10.1088/0004-637X/707/2/1334},
  \href {https://ui.adsabs.harvard.edu/abs/2009ApJ...707.1334W} {707, 1334}

\bibitem[\protect\citeauthoryear{{Zel'dovich} \& {Novikov}}{{Zel'dovich} \&
  {Novikov}}{1964}]{1964SPhD....9..246Z}
{Zel'dovich} Y.~B.,  {Novikov} I.~D.,  1964, Soviet Physics Doklady, \href
  {http://adsabs.harvard.edu/abs/1964SPhD....9..246Z} {9, 246}

\makeatother
\end{thebibliography}
\bibliographystyle{mnras} 


\label{lastpage}
\end{document}